\documentclass{aa}
\usepackage{graphics,latexsym,psfig}

\def\etal{{et\,al.}}
\def\msun{M$_{\odot}$}

\def\grad{$^\circ$}
\def\degs{\ifmmode ^{\circ}\else$^{\circ}$\fi}
\def\amin{\ifmmode ^{\prime}\else$^{\prime}$\fi}
\def\asec{\ifmmode ^{\prime\prime}\else$^{\prime\prime}$\fi}
\def\farcs{\hbox{$.\!\!^{\prime\prime}$}}  
\newbox\grsign \setbox\grsign=\hbox{$>$}
\newdimen\grdimen \grdimen=\ht\grsign
\newbox\laxbox \newbox\gaxbox
\setbox\gaxbox=\hbox{\raise.5ex\hbox{$>$}\llap
     {\lower.5ex\hbox{$\sim$}}}\ht1=\grdimen\dp1=0pt
\setbox\laxbox=\hbox{\raise.5ex\hbox{$<$}\llap
     {\lower.5ex\hbox{$\sim$}}}\ht2=\grdimen\dp2=0pt

\def\lax{$\mathrel{\copy\laxbox}$}

\def\h{$^{\rm h}$}\def\m{$^{\rm m}$}

\def\XMM{XMM-{\sl Newton}}

\begin{document}

\title{XMM-Newton observations of Nova LMC 2000\thanks{Partly based on
observations collected at the European Southern Obser\-vatory,
Chile under proposal 66.D-0391}}

\titlerunning{XMM-Newton observations of Nova LMC 2000}

\author{J. Greiner \inst{1}
     \and  M. Orio \inst{2,3}
     \and  N. Schartel \inst{4}
  }

  \offprints{J. Greiner, jcg@mpe.mpg.de}

\institute{
   Max-Planck-Institut f\"ur extraterrestrische Physik, 85741 Garching, Germany
   \and
   INAF - National Institute for Astrophysics,
   Osservatorio Astronomico di Torino, Strada Osservatorio 20, 
     10025 Pino Torinese, Italy
   \and
     Department of Astronomy, 475 N. Charter Str.,  Madison WI 53706, USA
   \and
     XMM-Newton Science Operations Centre, ESA,
     Villafranca del Castillo, P.O. Box 50727, 28080 Madrid, Spain
     }

\date{Received 27 February 2003 / Accepted 23 April 2003 }

\abstract{We report on three X-ray observations of Nova LMC 2000 with \XMM\
at 17, 51 and 294 days after the maximum, respectively. X-ray spectral fits
show a concordant decrease of the absorbing column and the X-ray luminosity.
No supersoft X-ray emission is detected. The mass of the ejected shell
is determined to be (less than) $7.5 \times 10^{-5}$ \msun.
Though data are sparse, one interesting correlation becomes visible:
sources with a long-duration supersoft X-ray phase have shorter
orbital periods than those with short or no supersoft X-ray phase.
This can be understood considering that (i) enough matter has to
be accreted in order to ignite the H burning, and (ii) that H burning
ceases when the mass of the remaining material (after shell ejection 
and burning) drops below a certain limit under which the temperature at
the bottom of the envelope is too low for the shell burning to compensate
the energy loss from the surface.
%
%
}

\maketitle

\section{Introduction}

Nova LMC 2000 was discovered in mid-July 2000 (Liller 2000).
Early observations between June 29.68 and
July 1.68 have detected the nova, but no brightness estimates are
available (Bond \& Kilmartin 2000), since all images except the first
one on June 29.68 show the nova to be saturated (Bond, priv. comm.).
This leaves the exact time and magnitude of the maximum unknown
except for the constraint that it was still rising on  June 29.68
(Bond, priv. comm.).
But even if the observed $m_{\rm V} = 11.2$ mag on 12 July 2000 was the 
maximum, with a corresponding to  $M_{\rm V} = -7.5$ mag, 
Nova LMC 2000 was among the 20\% brightest novae.

Optical spectroscopy on 15 July 2000 shows emission lines of the Balmer series,
several Fe multiplets and Na I D, suggesting a ``Fe II'' nova about one week 
after maximum (Duerbeck \& Pompei 2000). If this time estimate is correct,
the maximum of the nova could have been  $m_{\rm V} = 10.5$ mag (backwards
extrapolation to 8 July 2000) corresponding to  $M_{\rm V} = -8.2$ mag.

Spectroscopy of this ONeMg nova (Shore 2002) also revealed weak 
absorption components in the Balmer lines
at 1900 km/s on July 15 (Duerbeck \& Pompei 2000), and a FWZI of the
Balmer lines on July 14 and 15 of 2200$\pm$250 km/s 
(Hearnshaw \& Yan Tse 2000). Spectra obtained with 
HST on August 19/20, 2000, strongly resemble those of the fast nova V382 Vel
($t_3$ = 9 days; Della Valle \etal\ 2002)
at 2 months after visual maximum (Shore \etal\ 2000).

Only 4 out of about 100 classical novae observed with ROSAT 
have been found to exhibit a supersoft 
phase (see Orio \& Greiner 1999, Orio \etal\ 2002). 
However, theoretically one expects that each nova should pass
through a phase of soft X-ray emission during the later stages of decline
(e.g. MacDonald \etal\ 1985): During the post-maximum stage, at constant
bolometric luminosity, the photosphere progressively retreats as the 
residual hydrogen envelope is depleted, and the effective temperature
rises up to the soft X-ray region.
It is pretty much unclear yet what determines 
the appearance of supersoft X-ray emission.
It has been argued (Truran 2002) that the onset of the supersoft
phase is the same for all novae, about 6-8 months, and that
the duration of the supersoft X-ray phase is the burning timescale
of the white dwarf, i.e. proportional to the mass of the white dwarf
(Truran \& Glasner 1995, Vanlandingham \etal\ 2001).
While the first two supersoft novae, GQ Mus (\"Ogelman \etal\ 1993)
and V1974 Cyg (Krautter \etal\ 1996), are consistent
with this suggestion, observations of other recent novae imply that
the picture is more complicated.

\begin{table*}
\caption{Observation log \label{obslog}}
\begin{tabular}{ccccrcrr}
  \hline
  \noalign{\smallskip}
   Observation Interval & Time after   & Expo.-Time & Rev.  & Count rate &
                   HR1 & HR2~~ & HR3~~ \\
          (UT)      & Max. (days)  &  (ksec)~~  &       & (cts/ksec)\, & & & \\
 \noalign{\smallskip}
 \hline
 \noalign{\smallskip}
  2000-07-25 22:48 -- 2000-07-26 03:19 & ~17 & 16.26~~ & 115 & 5.0$\pm$0.6 &
       -- & 1.00$\pm$2.00 & 0.00$\pm$0.11  \\
  2000-08-28 13:53 -- 2000-08-28 16:40 & ~51 & 10.00~~ & 132 & 21.0$\pm$1.6 &
       0.70$\pm$0.07 & --0.56$\pm$0.06 & --0.66$\pm$0.16  \\
  2001-03-29 19:10 -- 2001-03-29 21:22 & 294 & 10.54~~ & 239 & $<$1.0~~~ &
      -- & --~~~~~~ & --~~~~~~ \\
  \noalign{\smallskip}
  \hline
  \noalign{\smallskip}
\end{tabular}

\noindent{(1) The hardness ratios are defined as 
              HR1 = $(B-A)/(B+A)$, HR2 = $(C-B)/(B+C)$ and HR3 = $(D-C)/(D+C)$,
           where $A (0.2-0.5$ keV), $B (0.5-2.0$ keV), $C (2.0-4.5$ keV),
           $D (4.5-7.5$ keV), 
            are the counts in the given energy range.}
\end{table*}

Here we report the results of a sequence of \XMM\ observations of
N LMC 2000 aimed at finding and characterizing the supersoft X-ray component.
We also report a contemporaneous  optical observation.

\section{Observations and Results}

\subsection{XMM-Newton observations}

Observations of Nova LMC 2000 were performed at three occasions 
(see Tab. \ref{obslog}), about 17, 51 and 294 days after the likely maximum on 
8 July 2000.

Throughout all observations the thin blocking filter was used.
In the following, we primarily deal with EPIC-pn 
(Str\"uder \etal 2001) part of the X-ray data,
and the optical monitor (OM;  Mason \etal\ 2001) data.

\begin{figure*}
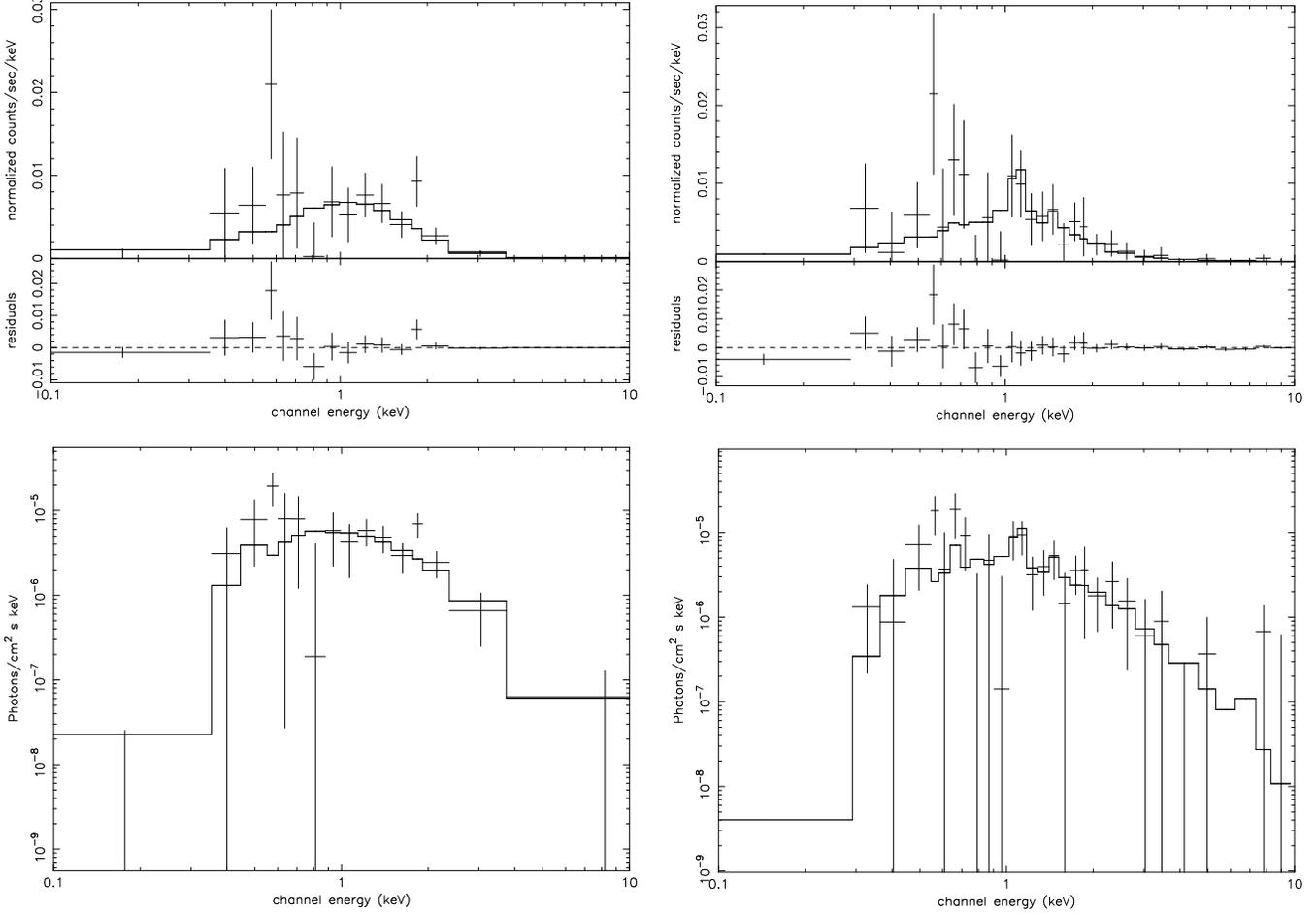

\vbox{\psfig{figure=nlmc00_c132sd_brems_csp.ps,width=8.8cm,angle=270,%
          bbllx=4.0cm,bblly=1.3cm,bburx=19.7cm,bbury=25.0cm,clip=}
\vspace{-5.8cm}\hspace{9.cm}
\psfig{figure=nlmc00_132sd_ray_csp.ps,width=8.8cm,angle=270,%
          bbllx=4.0cm,bblly=1.3cm,bburx=19.7cm,bbury=25.0cm,clip=}}
\medskip
\vbox{\psfig{figure=nlmc00_c132sd_brems_sp.ps,width=8.8cm,angle=270,%
          bbllx=2.8cm,bblly=1.2cm,bburx=20.1cm,bbury=25.cm,clip=}
\vspace{-6.35cm}\hspace{9.cm}
\psfig{figure=nlmc00_132sd_ray_sp.ps,width=8.8cm,angle=270,%
          bbllx=2.8cm,bblly=1.2cm,bburx=20.1cm,bbury=25.0cm,clip=}}
  \caption[spec]{X-ray spectrum of Nova LMC 2000 as observed with \XMM\
        on 28 August 2000, modelled with a bremsstrahlung (left) and
         Raymond-Smith model (right). The top panel shows the count rate
         spectrum and the residuals, while the lower panels show the
         unfolded photon spectrum.
             \label{spec}}
\end{figure*}

\subsubsection{XMM-Newton EPIC-pn data}

After a first inspection of the X-ray data, a source detection run was done
in the 0.5--7 keV band, as well as in narrow bands to 
derive hardness ratios. These hardness ratios, the count rates in the 
full band and the 3$\sigma$ upper limit for the non-detection during the
March 2001 observation are given in 
Tab. \ref{obslog} and show the strong intensity and spectral variability 
of the X-ray emission of Nova LMC 2000.

For the spectral analysis, single and double events away from the edges of the
CCD or bad pixels were extracted
in a source region of 19\asec, and a background region of outer diameter
of 30\asec.

Because of the larger number of collected photons we first fitted the 
X-ray spectrum obtained on 28 August 2000. Before the extraction 
of source and background photons, a time interval of $\sim$2 ksec was
excluded because of high background. Since the spectrum lacks any soft
component, we applied a bremsstrahlung and a Raymond-Smith model to the data.
Both models give satisfactory fits, and the fit parameters are given in
Tab. \ref{fitres}. 
During the observation on 28 August 2000, the unabsorbed X-ray luminosity
was 2$\times$10$^{34}$ erg/s (0.01--20 keV; assuming a distance of 55 kpc).
Note that despite the (best-fit) absorbing column of 
(2--3)$\times$10$^{21}$ cm$^{-2}$, a supersoft component with a characteristic
temperature of 30--50 eV and a luminosity of 10$^{36}$--10$^{38}$ erg/s
would have been easily detectable.

\begin{table*}
\caption{Spectral fit results of the July and August 2000 observations}
\vspace{-0.2cm}
\begin{tabular}{lccccc}
\hline
\noalign{\smallskip}
  Date   & $N_{\rm H}$ & $kT$ &  Normalization & Flux (bolometric) 
     & $\chi^{2}$ / dof \\
         & (10$^{22}$ cm$^{-2}$) & (keV) & (ph keV$^{-1}$ cm$^{-2}$s$^{-1}$)
            & (erg cm$^{-2}$s$^{-1}$)  & \\
\noalign{\smallskip}
\hline
\noalign{\smallskip}
                 & \multicolumn{5}{c}{bremsstrahlung model} \\
  25/26 Jul 2000 & 3.23$\pm$1.23 & 199$\pm$100 & 1.1E-05 & 7.5E-14 & 0.78 \\
  25/26 Jul 2000 & 7.00$\pm$2.57 & 1.6 fixed   & 3.6E-05 & 13.0E-14 & 1.28 \\
  28 Aug 2000    & 0.29$\pm$0.19 & 1.6$\pm$1.2 & 1.6E-05 & 5.5E-14 & 0.94 \\
\noalign{\medskip}
                 & \multicolumn{5}{c}{Raymond-Smith model}  \\
  25/26 Jul 2000 & 3.40$\pm$1.10 & 64$\pm$50.0 & 2.4E-05 & 6.9E-14 & 0.79 \\
  25/26 Jul 2000 & 7.65$\pm$2.33 & 2.1 fixed   & 7.5E-05 & 14.4E-14 & 1.00 \\
  28 Aug 2000    & 0.20$\pm$0.15 & 2.1$\pm$1.0 & 2.3E-05 & 4.5E-14 & 0.92 \\
\noalign{\smallskip}
\hline
\end{tabular}
\label{fitres}
\end{table*}

We then used the same models to also fit the July 25/26 observation.
However, due to the smaller number of counts, the temperature in both
models is not constrained. Thus, in a second iteration, we fixed the
temperature to the value as derived from the August 28, 2000 observation.
The best-fit parameters of both fits are also given in Tab. \ref{fitres}.
The unabsorbed bolometric luminosity of Nova LMC 2000 during the 
July 25/26, 2000 observation
was (5.0$\pm$0.3)$\times$10$^{34}$ (D/55 kpc)$^2$ erg/s, 
where the error includes also
the difference between the two models.
In a third iteration we tested a fit with the absorbing column fixed 
to the value of the July 2000 observation in order to check whether 
a temperature change could mimic the variation in absorbing column.
However, this does not provide an acceptable fit (reduced $\chi^2$ of 2.5).

For the X-ray non-detection on 29 March 2001, assuming the same 
spectral models and no intrinsic absorption through the ejected nova shell
we derive a 3$\sigma$ upper limit for the luminosity of
$L < 1.1 \times 10^{33}$ (D/55 kpc)$^2$ erg/s for the case of only
galactic foreground absorption (7$\times$10$^{20}$ cm$^{-2}$) or 
$L < 1.4 \times 10^{33}$ (D/55 kpc)$^2$ erg/s for the case of 
galactic foreground plus total LMC absorption 
(15$\times$10$^{20}$ cm$^{-2}$; Luks 1994).

\subsubsection{XMM-Newton Optical Monitor data}

The OM was used with a variety of filters, and also the grism was used
in two occasions. In particular, we obtained
(i) two 5 ksec grism 2 (visual spectrum) exposures, a sequence of
  5 exposures with the UVW1 filter, and a sequence of 4 exposures with the
  UVW2 filter, each with 940 sec) in the first  XMM-Newton observation
  of Nova LMC 2000 (revolution 115);
(ii)  a sequence of  5 exposures with the V band 
 filter with 1 ksec each, followed by a 5 ksec grism 2 exposure 
  (second XMM-Newton observation; rev. 132);
(iii) a sequence of 5 exposures of 1 ksec with the U band filter
  in the last \XMM\ observation (rev. 239).

Unfortunately, the grism data are hardly usable since the window of the
CCD was set too small. The photometry in the different filters is
summarized in Tab. \ref{om}. The central wavelengths for the UV filters
are 2910 \AA\ for the UVW1 and 2120 \AA\ for the UVW2, respectively.

\begin{table}[ht]
\caption{XMM/OM photometry \label{om}}
\begin{tabular}{ccccc}
  \hline
  \noalign{\smallskip}
  Date & filter & $\!\!$countrate$\!\!$ & brightness$^{(1)}$ & flux$^{(1)}$ \\
              &         & cts/sec  & mag & erg/cm$^2$/s/\AA \\
  \noalign{\smallskip}
  \hline
  \noalign{\smallskip}
    $\!\!$2000-07-25 & UVW1 & 201.4    & 11.41$\pm$0.01 & 1.04E-13 \\
                     & UVW2 & ~16.0    & 11.62$\pm$0.01 & 1.24E-13 \\
    $\!\!$2000-08-28 & V    & ~10.6    & 15.24$\pm$0.02 & 3.05E-15 \\
    $\!\!$2001-03-29 & U    & ~~9.5    & 15.70$\pm$0.02 & 2.24E-15 \\
  \noalign{\smallskip}
  \hline
  \noalign{\smallskip}
\end{tabular}

\noindent{(1) Brightness and flux are given without correction for 
  galactic, LMC and nova-intrinsic extinction, since the latter two
  values have uncertainties much larger than the measurement error.}
\end{table}

\subsection{Optical photometry}

\begin{figure*}
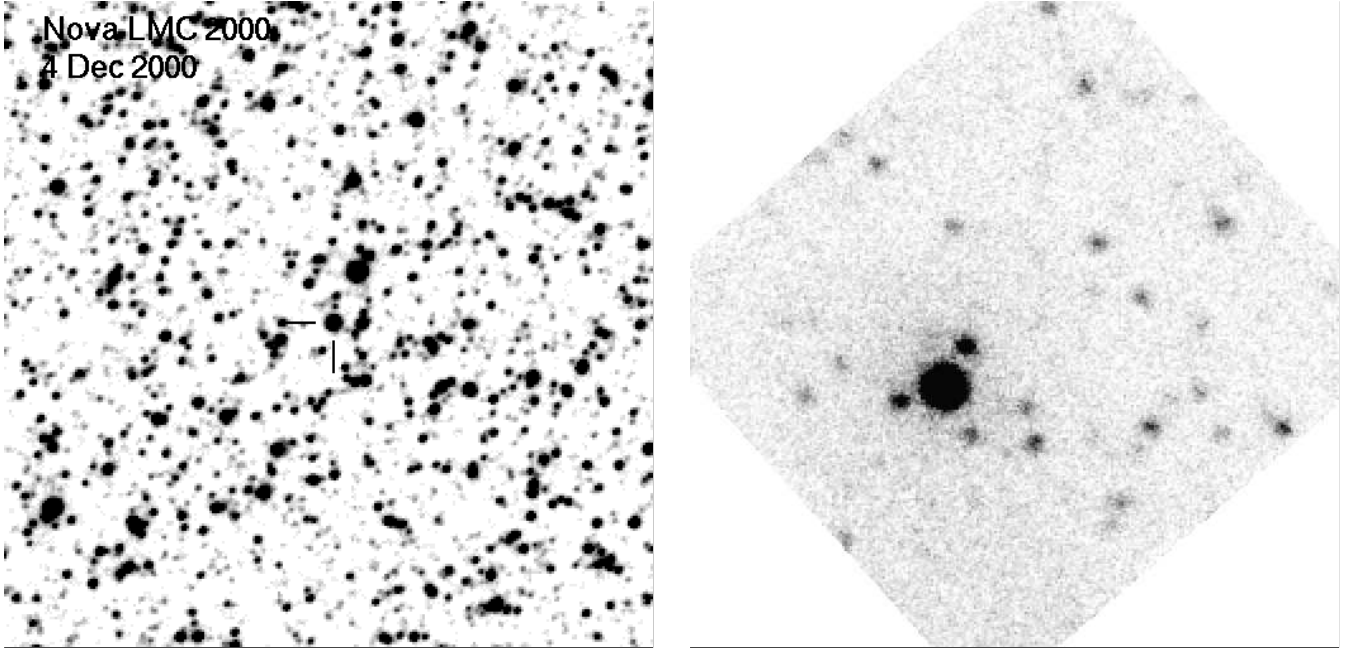

  \begin{tabular}{cc}
  \vbox{\psfig{figure=nlmc2000_fc.ps,width=8.7cm,%
         bbllx=3.05cm,bblly=10.5cm,bburx=18.7cm,bbury=26.2cm,clip=}}\par &
  \vbox{\psfig{figure=nlmc2000_000725_uvw2.ps,width=8.7cm,%
         bbllx=3.05cm,bblly=10.5cm,bburx=18.7cm,bbury=26.2cm,clip=}}\par \\
  \end{tabular}
 \caption[fchart]{Finding chart of N LMC 2000 in the V band (left; marked 
    with two dashes near the image center) and at
    2120 \aa\ (right). The V band image was taken on 2000 Dec. 4
    with DFOSC at the 1.5\,m Danish telescope at La Silla (ESO).
    We measure RA = 5\h25\m01\fs1 and Decl. = -70\grad 14\amin 17\asec\
    (equinox 2000.0) with an error radius of 2\asec.
    The field size is 1\farcm7$\times$1\farcm7.
    The right image is the sum of 4 images with 1000 sec exposure each, 
    taken by the OM onboard XMM in the UVW2 filter on 25/26 July 2000,
    where the nova is the by far brightest object in the image.
     The field size is 2\farcm3$\times$2\farcm3.
    North is up and East to the left.}
       \label{fc}
 \end{figure*}

\begin{figure}[ht]
 \psfig{figure=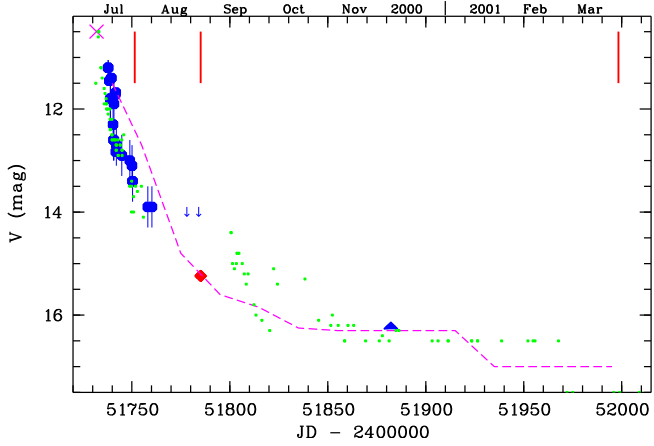,width=8.8cm,%
          bbllx=1.9cm,bblly=1.3cm,bburx=19.3cm,bbury=12.9cm,clip=}\par
  \caption[lcurve]{Light curve of Nova LMC 2000 as 
      measured by various observers
       and provided to VSNET (filled circles). Our ground-based observation 
       from Dec. 2000
       is marked by a filled triangle, and the XMM/OM observation in
       August 2000 by a filled lozenge. The cross marks the possible maximum 
       as determined by a backwards extrapolation (see Introduction). 
       Vertical dashes at the top mark the times 
       of the \XMM\ observations. The dashed line visualizes the decay
       curve of the recurrent nova CI Aql (see Greiner \& Di\,Stefano 2002),
       scaled to the LMC distance. While the early decay of N LMC 2000
       was substantially faster, the optical brightness in Dec. 2000 is 
       consistent with the nova still being in the plateau phase.
       For comparison purposes also overplotted (with small dots) 
       is the VSNET light curve of
       nova V1494 Aql (scaled in brightness) which has very similar
       $t_2$ and $t_3$ times. }
       \label{lcur}
    \end{figure}

On 4 December 2000 we obtained three V band exposures with 90 sec exposure time
each using DFOSC at the 1.5\,m Danish telescope at La Silla (ESO, Chile).
The MAT/EEV 44-82 CCD chip with 15 $\mu$m pixels has a plate scale of
0\farcs4/pixel. The seeing was 1\asec. 
The photometric standard RU 149, observed at nearly the same airmass as
N LMC 2000, was used for the absolute flux calibration.
Due to the brightness of the nova and the relatively crowding-free surrounding
(see Fig. \ref{fc}),
aperture photometry was done within the MIDAS package after standard
flatfield and bias correction.
We measure V = 16.20$\pm$0.03 mag independently in each of the three images 
for  Nova LMC 2000.
After adding this to the early measurements by VSNET observers 
(Fig. \ref{lcur}),
a comparison with light curve models of Hachisu \& Kato suggests that 
Nova LMC 2000 was possibly still in the plateau phase during our optical
observation on 4 December 2000. 

A crude estimate of the decay rate based on the VSNET data gives 
$t_2$ $\sim$ 9.0 days and $t_3$ $\sim$ 22 days, with an error of 
$\pm$2 days depending on the actual occurrence of 
the not observed optical maximum 
(where $t_2$ and $t_3$ are the times 
during which the brightness decays by two and three magnitudes, respectively).

\section{Discussion}

\subsection{Luminosity and absorption change}

The 0.1--10 keV X-ray emission observed from Nova LMC 2000 at 17 and 51 days
after the explosion is well described by a bremsstrahlung or 
Raymond-Smith model. Spectral fitting reveals that during the first
observation the absorbing column was much larger, indicating
intrinsic absorption due to the ejected shell. This is independent of
the model used in the fitting. Despite the
smaller count rate in the first observation with respect to the second,
the derived luminosity was largest in the first observation.

\subsection{The mass of the ejected envelope}

The large excess absorption determined from the first observation can be used
to estimate the mass of the ejected shell, assuming no clumpiness
(and a filling factor of 1 though this is likely an overestimate), 
i.e. that the measured column density is 
representative for all spatial directions around the nova system.
At the time of the first observation, the expansion velocity was
$\sim$2000 km/s (Duerbeck \& Pompei 2000, Hearnshaw \& Yan Tse 2000),
and thus the shell radius was $R \sim 3\times 10^9$ km, if no substantial
deceleration has occurred during the first 17 days after the nova explosion.
Using a mean column density of neutral hydrogen of $5\times 10^{22}$ cm$^{-2}$
(see Tab. \ref{fitres}), a mass of the shell of 
$\Delta M_{ej} = 7.5 \times 10^{-5}$ \msun\ is obtained.
As a kind of consistency check one can also consider the later decrease
of the absorbing column between the first and second \XMM\ observation.
Since the time difference, and thus the radius difference,
is a factor of 3, the absorbing column should
have reduced by a factor of 9 if the shell had expanded with constant velocity.
This is, within the errors, consistent with the measured change 
in column density (see Tab. \ref{fitres}).

An independent estimate of the ejected shell mass can be derived from the
measured decay time $t_2$ and the relation 
$log\ \Delta M_{ej} = 0.274 \times {\rm log}\ t_2 - 4.355$
(Della Valle \etal\ 2002; note that their formula as well as their
figure miss a factor of 10$^{-5}$ in  $\Delta M_{ej}$). This yields 
$\Delta M_{ej} = 8 \times 10^{-5}$ \msun, in good agreement
with the above estimate from the measured column density (note though that the 
Della Valle \etal\ relation refers to the ionized hydrogen mass).

Based on the relation between decay time vs. absolute magnitude  
(Della Valle \& Livio
1995) and our measured $t_2$ value, the absolute magnitude of Nova LMC 2000
would be $M_{\rm V} = -8.74$ mag, half a magnitude brighter than
the backwards extrapolation based on the spectral appearance. This 
would imply that this extrapolation is reasonable.

\subsection{The supersoft X-ray phase in novae}

No supersoft X-ray phase has been found for Nova LMC 2000: while on 
25/26 July 2000 the ejected shell was still too optically thick to allow soft
X-rays to penetrate, the shell had thinned considerably over the next
4 weeks (until Aug. 28) to allow the detection of supersoft emission, 
if it had been present.
Thus, the supersoft X-ray phase in Nova LMC 2000 must have been
shorter than 7 weeks, or 6$\times t_2$.
In view of the fact that an HST STIS spectrum taken Aug. 19/20, i.e.
9 days before the second \XMM\ observation shows a 1150--3120 \AA\ flux of
1.6$\times$10$^{38}$ erg/s (Shore \etal\ 2000), the shell burning should 
either has switched
off during those 9 days, or the effective temperature was below $\sim$10 eV
in order not to be observable in our \XMM\ exposure.

In an attempt to understand the reason for the largely varying duration
of the supersoft X-ray phase in different novae we have compiled some
information on novae with observed supersoft X-ray phase or very
constraining observational limits (Tab. \ref{ssnovae}).

 \begin{figure}[ht]
 \psfig{figure=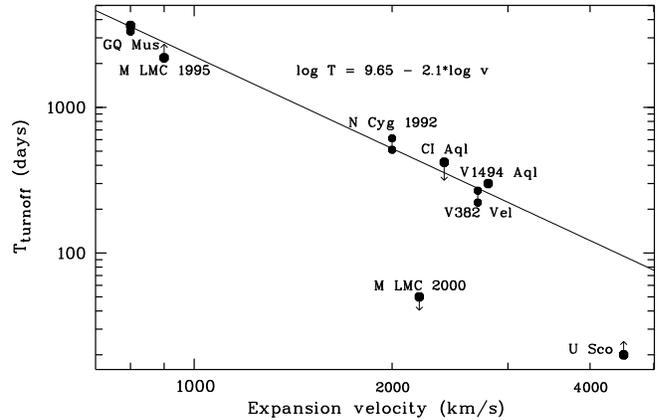,width=8.8cm,%
          bbllx=1.3cm,bblly=1.1cm,bburx=19.3cm,bbury=12.9cm,clip=}\par
  \caption[turnoff]{Correlation between the envelope expansion velocity and the
      turn-off times of novae (see Tab. \ref{ssnovae}).
      For GQ Mus, Nova Cyg 1992 and V382 Vel two points are plotted, 
      resembling the temporal limits between observed supersoft X-ray emission
      and H burning switch-off. For CI Aql and U Sco only limits are
      available. Nova LMC 2000 does not fit into this correlation unless
      the intrinsic absorbing column for the supersoft X-ray emission
      has been muc larger than estimated in Sect. 2 and 3. 
      If the relation were true, one would predict that IM Nor with its
      rather small expansion velocity would turn into a long-lived 
      ($\sim$3 yrs) supersoft X-ray phase.
       \label{vel}}
    \end{figure}

\begin{table*}
\caption{Details on novae with detected supersoft X-ray emission (top 5 lines),
 N LMC 2000, and recent recurrent novae (RN; last three lines). 
 \label{ssnovae}}
\vspace{-0.2cm}
\begin{tabular}{lcccccccll}
  \hline
  \noalign{\smallskip}
    Nova Name &   $t_2$ & $t_3$ & $M_{\rm WD}$ & log $\Delta M_{ej}$ & 
       $V_{\rm expa}$$^{(c)}$ & $P_{\rm orb}$  &  log $\Delta M_{ign}$
         &  Limits on supersoft  phase & Refs. \\
         & (days) & (days)& (\msun)$^{(a)}$ & $^{(b)}$  & (km/s)& (days) && & $^{(d)}$ \\
 \noalign{\smallskip}
 \hline
 \noalign{\smallskip}
   GQ Mus      & 23.0  &  45  & 0.87 & -3.98 & 800 & 0.0593 & -3.4 & 10 yrs 
   & 1--3 \\
   N Cyg 1992  & 16.0  &  42  & 0.90 & -4.02  & 2000  & 0.0812 & -3.8 & turnoff at days 32--38$\times t_2$ & 4--6 \\
   N LMC 1995  &  12$^{(e)}$ & 16$^{(e)}$ & 1.23 & -4.22 & ~\,900 & -- & & $>$6 yrs and continuing & 7--9 \\
   V382 Vel    &  4.5  &   9  & 1.17 & -4.18 & 2700 & 0.1461 & -4.8 &at 44$\times t_2$; 
       none at 60$\times t_2$ & 10 \\
   V1494 Aql   &  ~~~6.6$^{(f)}$ & ~~~16$^{(f)}$  & 1.09 &  -4.13 & 2800 &  0.1346 &
     -4.6 &  none \lax 28$\times t_2$, but at 38--45$\times t_2$  & 11, 12 \\
 \noalign{\medskip}
   N LMC 2000  &  9.0  &  22  & 1.02 & -4.09  & 2200 & -- & & less than 5.5$\times t_2$ & \\
 \noalign{\medskip}
   U Sco (RN)  & 5.0 & 7 & 1.20 &  ~~~-5.99$^{(g)}$ &4500 & 1.230 & -5.3 & at 4$\times t_2$ & 13 \\
   IM Nor (RN) & ~~~20.5$^{(b)}$ &  ~~~47$^{(b)}$  & 0.85 & -3.99   & 1150 & -- & & 
      none at 1.1$\times t_2$ & 12, 14 \\
   CI Aql (RN) & 30.0 &  36  & 0.94 & -3.95  &2400  & 0.6184 & -5.2 & less than 14$\times t_2$ & 15, 16 \\
  \noalign{\smallskip}
  \hline
  \noalign{\smallskip}
\end{tabular}

\noindent{(a) According to the $t_3-M_{\rm WD}$ relation (Livio 1992) \\ 
          (b) According to  
              log $\Delta M_{ej}$ = 0.274 $\times$ log $t_2$ - 4.355 
              (Della Valle \etal\ 2002).\\
          (c) Usually derived for the H$\alpha$ line. For V382 Vel, lacking
               an estimate based on H$\alpha$, we used the number given for 
               UV lines and divide by a factor 2 (Shore 2002). \\
          (d) References: 
             (1) \"Ogelman \etal\ 1993,
             (2) Pequignot \etal\ 1993,
             (3) Shanley \etal\ 1995,
             (4) Krautter \etal\ 1996,
             (5) Chochol \etal\ 1993,
             (6) Balman \etal\ 1998,
             (7) Della Valle \etal\ 1995,
             (8) Orio \& Greiner 1999, 
             (9) Orio \etal\ 1993,
            (10) Orio \etal\ 2002,
            (11) Venturini \etal\ 2000,
            (12) Starrfield  \etal\ 2002,
            (13) Iijima 2002,
            (14) Duerbeck \etal\ 2002,
            (15) Greiner \& Di\,Stefano 2002,
            (16) Kiss \etal\ 2001. \\
           (d) Determined here from the VSNET light curves. \\
           (g) Instead of the prediction we used the value as measured
               by Della Valle \etal\ (2002).
           }
\end{table*}

The detectability of supersoft X-ray emission from novae depends on several
factors, the three most important ones (probably) being the existence/duration
of such a supersoft X-ray phase in the first place, the amount of
matter ejected during the thermonuclear runaway and that blown away later 
during a wind phase, both of
which hide the supersoft X-ray emission during the early 
phase after a nova until the shell has become optically thin.
It is therefore interesting to ask how these factors relate to
other observable parameters of a nova?

The ignition of the thermonuclear runaway of accreted hydrogen is
primarily determined by the pressure at the base of the accreted
envelope, which in turn depends on the mass and radius of the white dwarf
and the mass of the envelope (Fujimoto 1982a).
It has been argued that the $t_3$-time 
primarily depends on the mass of the white dwarf (eq. 12 in Livio 1992).
Furthermore, since also the maximum absolute magnitude of a nova correlates
with the mass of the white dwarf as well as the decline rate, Della Valle
\etal\ (2002) have recently found a correlation between the shell mass and
the  $t_2$-time (note that Shore 2002 has proposed an alternative relation
using the same quantities). 
If the duration of the supersoft X-ray phase depended 
on the amount of mass which is left over after the ejection of the shell, i.e.
the difference between the mass necessary to ignite the thermonuclear
explosion and the mass of the matter ejected into the expanding shell,
it could be expected that the supersoft X-ray phase would correlate
with the $t_2$- or  $t_3$-time. We have collected the relevant data (see
Tab. \ref{ssnovae})
for the five novae with detected supersoft X-ray emission, plus a few
cases with stringent limits (including the present case Nova LMC 2000).
Comparing the decay times (columns 2 and 3) and the implied
envelope mass (col. 4) according to Della Valle \etal\ (2002) with the 
duration of the supersoft X-ray phase (col. 8) does not show any obvious
correlation, however (though we note that envelope mass estimates are
typically very uncertain). Thus, the supersoft X-ray phase is 
seemingly not directly correlated to 
the brightness decay rate and/or the mass of the white dwarf.
This also argues against the simple scenario of Truran \& Glasner (1995)
and Vanlandingham \etal\ (2001):
for instance, N LMC 1995 with a rather short decline time and a 
 white dwarf mass of $\sim$1.2 \msun\
according to the $t_3$-$M_{\rm WD}$ relation would have an expected
supersoft X-ray phase of less than 1 yr, but is observed as supersoft
X-ray source for over 6 years now (Orio \etal\ 2003).

We have also considered the expansion velocities of the nova shells
(col. 6 in Tab. \ref{ssnovae}), since it is proportional to the pressure 
at the base of the accreted shell. There is an interesting
correlation with the duration of the supersoft X-ray phase in 8 sources
except for Nova LMC 2000 which falls off completely (Fig. \ref{vel}).
Thus, either this correlation is chance coincidence due to the small
number statistics, or the non-detection the supersoft X-ray phase
is caused by a much higher than estimated absorbing column for which we
have no other evidence. One could imagine that the (hard) X-ray
emission which is seen on 28 August 2000 originates outside the
expelled envelope, and therefore the small absorbing column derived
from the fit of that emission is not representative of the column
which blocks the supersoft emission from the white dwarf. 
A third alternative is that the burning continued much longer, but 
at such low temperature ($<$10 eV) that made it impossible to be detected
by \XMM.

\begin{figure}[ht]
 \psfig{figure=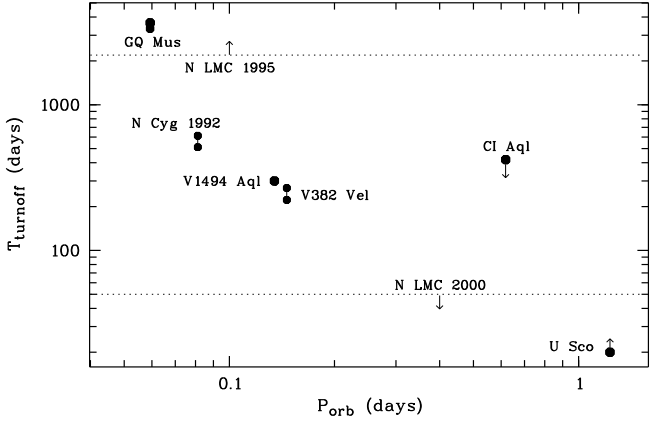,width=8.8cm,%
          bbllx=1.3cm,bblly=1.1cm,bburx=19.3cm,bbury=12.9cm,clip=}\par
  \caption[turnoff]{Correlation between the orbital periods and the
      turn-off times of novae (see Tab. \ref{ssnovae}).
       Symbols and limits as in Fig. \ref{vel}.
       \label{toff}}
    \end{figure}

A somewhat surprising correlation is found when plotting the orbital period
over the X-ray turn-off time (Fig. \ref{toff}): systems with a short orbital
period have a long H shell burning period. One possible explanation would
be the higher irradiation of the companion star in short orbital period
binaries, which in turn may increase the mass transfer rate to the 
white dwarf. But even without considering irradiation, the orbital period is
strongly correlated to the mass transfer rate in cataclysmic binaries
(e.g. Patterson 1984), and therefore can be expected to be a dominant
factor in the evolution of a nova through a supersoft X-ray phase (see below).

 \begin{figure}[ht]
 \psfig{figure=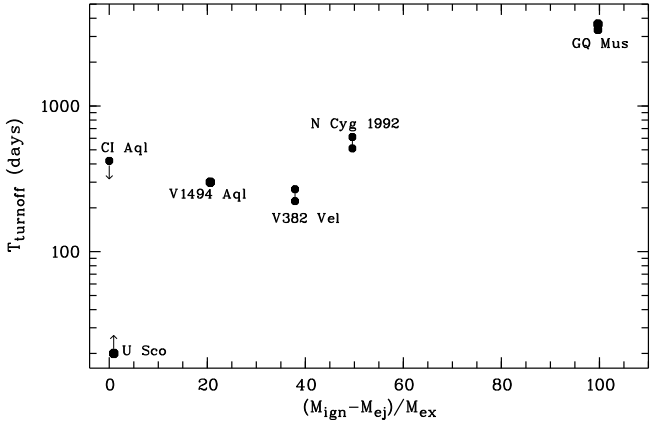,width=8.8cm,%
          bbllx=1.3cm,bblly=1.1cm,bburx=19.3cm,bbury=12.9cm,clip=}\par
  \caption[turnoff]{Correlation between the 
      turn-off times of novae (see Tab. \ref{ssnovae}) and the ratio
      of left-over mass to the critical mass where H shell burning will cease.
      Symbols and limits as in Fig. \ref{vel}.
       \label{mass}}
    \end{figure}

A  different way to consider the problem of the supersoft X-ray phase
duration is to remember
that the H shell burning may cease before all material
is consumed. As has first been argued by Fujimoto (1982b), burning
will cease when the mass of the shell becomes less than a critical
value $M_{\rm ext}$, below which the temperature at
the bottom of the envelope is too low for the shell burning to compensate
the energy loss from the surface. Thus, the important factor is not the
absolute value of the matter left after the shell has been expelled, but the 
difference of this left-over matter and the critical mass $M_{\rm ext}$.
In evaluating this suggested correlation (Fujimoto 1982b), we have 
(i) derived the accretion rate from the orbital period using the
  ``master equation of binary evolution'' for the case of rotational braking
   (eq. 38 in Patterson 1984),
(ii) deduced the white dwarf mass from the $t_3$ time (eq. 13 in Livio 1992),
(iii) by using the accretion rate and white dwarf masses
   determined the mass ($\Delta M_{\rm ign}$) necessary to ignite the 
   shell burning
    depending on the white dwarf mass and accretion rate (Fig. 7 from Fujimoto
   1982b;  see column 8 in Tab. \ref{ssnovae}),
(iv) derived the ejected shell mass from the $t_2$ time (Della Valle 
   \etal\ 2002; see column 5 in Tab. \ref{ssnovae}),
(v) compared the difference of the total shell mass ($\Delta M_{\rm ign}$)
  and the ejected mass ($\Delta M_{\rm ej}$) with the critical mass 
  ($\Delta M_{\rm ex}$) which varies between $\sim 5 \times$10$^{-5}$ \msun\ 
  for a 0.8 \msun\ white dwarf mass down to $\sim$10$^{-6}$ \msun\ for
  a 1.2 \msun\ white dwarf mass. 

Fig. \ref{mass} shows this ratio in dependence of the X-ray turn-off time
for the six novae for which the relevant data ($t_2$, $t_3$, $P_{\rm orb}$)
are available.
We find that three sources have substantial mass above the critical
$\Delta M_{\rm ex}$: GQ Mus has by far the largest mass excess, consistent 
with its
$\sim$10 yr supersoft X-ray phase, as well as N Cyg 1992 and V1494 Aql.
The other three systems, V382 Vel, CI Aql and U Sco happen to fall
very near their $\Delta M_{\rm ex}$ values, and according to the
suggestion of Fujimoto (1982b) would turn off very rapidly after
the explosion. This is consistent with the observations.

Of course, there are several limitations to this simple picture,
and additional factors are thought to also influence the duration of 
the supersoft X-ray phase, but were ignored here:
(i) Metallicity is certainly an important factor which determines the 
  appearance of supersoft X-rays. With Nova LMC 2000 having the same
  metallicity and additionally a very similar environment (less than 
  20 arcmin offset) as the supersoft Nova LMC 1995 (Orio \& Greiner 1999),
  one could have hoped for a similar X-ray behaviour. But obviously
  the limits on the supersoft phase in Nova LMC 2000 imply a much
  shorter duration than the more than 6 years (and continuing)
  of Nova LMC 1995 (Orio \etal\ 2003).
(ii) It is generally agreed upon that there is a continuous, 
  long-lasting loss of matter (wind)
  after the possible ejection at the time of the thermonuclear runaway. 
  Wind-driven mass loss is expected
  to be strongly mass dependent, i.e. the
  radiation pressure grows for larger white dwarf mass. This would imply
  a shorter supersoft X-ray phase for novae with larger  white dwarf masses
  (e.g. Starrfield \etal\ 1991, Yungelson \etal\ 1996).

Unfortunately, no statement can be made concerning  N LMC 2000.
It would be interesting to determine the orbital periods of N LMC 2000
and N LMC 1995 to determine their location relative to $\Delta M_{\rm ex}$,
and thus to test the above hypothesis that the duration of the supersoft
X-ray phase is determined by the ratio of the left-over mass to the
critical mass $\Delta M_{\rm ex}$ for shell burning.
If the above relations hold, we would infer orbital periods in the range
of 0.5--1 day for Nova LMC 2000 and 2 hrs for Nova LMC 1995.

\section*{Acknowledgments}
We are highly indebted to F. Jansen for granting \XMM\ Director's 
discretionary time for these Target of Opportunity observations
of Nova LMC 2000.
We are grateful to the VSNET observers for providing most of the measurements
plotted in Fig. 3.
JG thanks I. Bond for the details of the early photometry obtained within
the MOA project (www.vuw.ac.nz/scps/moa), as well as C. James and
A. Breeveld for the help in the attempt to make use of the OM grism data.
Based on observations obtained with XMM-Newton, an ESA science
  mission with  instruments and contributions directly funded by
  ESA Member States and NASA.


\begin{thebibliography}{}

\bibitem[]{bog98} Balman S., Krautter J., \"Ogelman H., 1998, ApJ 499, 395

\bibitem[]{bk00} Bond I.A., Kilmartin P.M., 2000, IAUC 7457

\bibitem[]{chu93} Chochol D., Hric L., Urban Z., \etal, 1993, A\&A 277, 103

\bibitem[]{dp00} Duerbeck H.W., Pompei E., 2000, IAUC 7457

\bibitem[]{dbd02} Duerbeck H.W., Baptista R., Dutra C.M., Sterken C., 2002,
  IAUC 7799

\bibitem[]{dvl95} Della Valle M., Livio M., 1995, ApJ 452, 704

\bibitem[]{dmb95} Della Valle M., Masetti N., Benetti S., 1995, IAUC 6144

\bibitem[]{dvp02} Della Valle M., Pasquini L., Daou D., Williams R.E., 2002,
   A\&A 390, 155

\bibitem[]{fuj82a} Fujimoto M.Y., 1982a, ApJ 257, 752

\bibitem[]{fuj82b} Fujimoto M.Y., 1982b, ApJ 257, 767

\bibitem[]{gds02} Greiner J., DiStefano R., 2002, ApJ 578, L59

\bibitem[]{ht00} Hearnshaw J.B., Yan Tse J.,  2000, IAUC 7457

\bibitem[]{iji2} Iijima T., 2002, A\&A 387, 1013

\bibitem{kto01} Kiss L.L., Thomson J.R., Ogloza W., F\"uresz G.,
  Sziladi K., 2001, A\&A 366, 858

\bibitem[1996]{Krautter}  Krautter J., \"Ogelman  H., Starrfield
S., Wichmann R., Pfeffermann E., 1996, ApJ 456, 788

\bibitem[]{Lil00} Liller W., 2000, IAUC 7453

\bibitem[]{Liv92} Livio M., 1992, ApJ 393, 516

\bibitem[]{Luk94} Luks T., 1994, RvMA 7, 171

\bibitem[]{mft85} MacDonald J., Fujimoto M.Y., Truran J.W., 1985, ApJ 294, 263

\bibitem[]{mbm01} Mason K.O., Breeveld A., Much R. \etal\ 2001,
    A\&A 365, L36

\bibitem{ooks93} \"Ogelman H.,  Orio M., Krautter J., Starrfield S., 1993,
   Nat. 361, 331

\bibitem[]{org99} Orio M., Greiner J., 1999, A\&A 344, L13

\bibitem[]{ori02} Orio M., Parmar A.N., Greiner J., \etal\ 2002, MN 333, L11 

\bibitem[]{ori03} Orio M., Hartmann W., Still M., Greiner J., 2003, 
  ApJ (submitted)

\bibitem[]{patt84} Patterson J., 1984, ApJS 54, 443

\bibitem[]{ppb93} Pequignot D., Petitjean P., Boisson C., Krautter J., 1993,
  A\&A 271, 219

\bibitem[]{sog95} Shanley L., \"Ogelman H., Gallagher J.S., Orio M., 
   Krautter J., 1995, ApJ 438, L95

\bibitem[]{ssb00} Shore S.N., Starrfield S., Bond H.E., Downes R., 
 Hauschildt P.H., Gehrz R.D., Woodward C.E., Krautter J., Evans A.N., 2000,
IAUC 7486

\bibitem[]{ssn02} Shore S.N., 2002, in Classical Nova Explosions,
Eds. M. Hernanz \& J. Jos\'e, AIP Conf. Proc. 637, p. 175


\bibitem[]{sts91} Starrfield S., Truran J.W., Sparks W.M., Krautter J., 1991,
  in Extreme Ultraviolet Astronomy, ed. R.F. Malina \& S. Bowyer (New York:
  Pergamon), p. 168

\bibitem[]{sta02} Starrfield S., 2002, in Classical Nova Explosions,
  Eds. M. Hernanz \& J. Jos\'e, AIP Conf. Proc. 637, p. 89

\bibitem[]{sbd01} Str\"uder L., Briel U., Dennerl K., \etal\ 2001, 
   A\&A 365, L18

\bibitem[]{tru02} Truran J.W., 2002, in 
  The Physics of Cataclysmic Cariables and Related Objects, eds.  
  B.T. G\"ansicke, K. Beuermann, K. Reinsch, ASP Conf. 261, p. 576

\bibitem[]{tg95} Truran J.W., Glasner S.A., 1995, in Cataclysmic Variables,
  eds. A. Bianchini, M. Della Valle, M. Orio, ASSL 205, 453

\bibitem[]{vss01} Vanlandingham K.M., Schwarz G.J., Shore S.N., Starrfield S.,
  2001, AJ 121, 1126 

\bibitem[]{vrl00} Venturini C., Rudy R.J., Lynch D.K., Mazuk S., 
  Puetter R.C., Armstrong T., 2000, IAUC 7490

\bibitem[1996]{Yungelson} Yungelson L., Livio M., Truran J.W., \etal, 1996,
   ApJ 466, 890



\end{thebibliography}
\end{document}